\documentclass{ws-mpla}
\usepackage{amssymb}

\usepackage[super]{cite}
\usepackage{graphicx}

\begin{document}

\title{Local Varying-Alpha Theories}
\author{John D. Barrow \and Jo\~{a}o Magueijo}
\maketitle

\begin{abstract}
In a recent paper we demonstrated how the simplest model for varying alpha
may be interpreted as the effect of a dielectric material, generalized to be
consistent with Lorentz invariance. Unlike normal dielectrics, such a medium
cannot change the speed of light, and its dynamics obey a Klein-Gordon
equation. This work immediately suggests an extension of the standard
theory, even if we require compliance with Lorentz invariance. Instead of a
wave equation, the dynamics may satisfy a local algebraic relation involving
the permittivity and the properties of the electromagnetic field, in analogy
with more conventional dielectric (but still preserving Lorentz invariance).
We develop the formalism for such theories and investigate some
phenomenological implications. The problem of the divergence of the
classical self-energy can be solved, or at least softened, in this
framework. Some interesting new cosmological solutions for the very early
universe are found, including the possibility of a bounce, inflation and
expansion with a loitering phase, all of which are induced by early
variations in alpha.
\end{abstract}


\markboth{John D. Barrow and Joao Magueijo}
{Local Varying-Alpha Theories}

\catchline{}{}{}{}{} 

\address{DAMTP, Centre for Mathematical Sciences, University of Cambridge,
Wilberforce Rd., Cambridge, CB3 0WA, U.K.}

\address{Theoretical Physics, Blackett Laboratory, Imperial College, London, SW7 2BZ,
U.K.}

\pub{Received (Day Month Year)}{Revised (Day Month Year)}

\ccode{PACS Nos.: 98.80.-k, 06.20.Jr, 04.20.-q}

\section{Introduction}

Since 1999 there have been a series of fascinating high-precision
observational investigations into the possible space and time variation of
some of the traditional constants of physics, notably the fine structure
constant, $\alpha \equiv e^{2}/\hslash c$ and the proton-electron mass ratio 
$\mu \equiv m_{p}/m_{e}$, (and combinations of the two) which have been
extensively reviewed \cite{con}. Indications of possible variations of $%
\alpha $\ in time~\cite{webb1999} and space \cite{webb2011} and time
variations in $\mu $~\cite{Reinhold2006} have been reported. These
observations have inspired the creation of self-consistent theories in which
'constants' like $\alpha $ and $\mu $ are promoted to become scalar fields
that gravitate with their own dynamics that conserve energy and momentum 
\cite{sandvik2002}. Such theories extend the philosophy of Jordan \cite%
{jordan} and Brans-Dicke \cite{bd}, who first created extensions of general
relativity to accommodate variations in the Newtonian gravitation constant $%
G $, to variations of other non-gravitational \textquotedblleft
constants\textquotedblright .

Variations in the traditional (low-energy) constants of physics offer a new
observational window on fundamental physics at very high-energies. Theories
with a non-unique vacuum state, possessing \textquotedblleft
extra\textquotedblright\ dimensions, or permitting new light scalar fields,
can all lead to space-time variations of the fundamental low-energy
\textquotedblleft constants\textquotedblright\ of nature \cite%
{Bekenstein1982, con}. Simultaneous variations of different gauge couplings
may be significantly constrained by nay form of grand unification at
sub-Planckian energies \cite{drink}.

Small variations of non-gravitational constants have negligible effects on
the expansion dynamics of the universe but have potentially observable
influences on astronomical spectra from atomic and molecular transitions 
\cite{sav}. Self-consistent scalar-tensor theories are needed to evaluate
their full cosmological consequences. Theoretical studies have focused on a
varying fine-structure constant $\alpha $, which is simplest to develop
because of its gauge symmetry \cite{sandvik2002}, and a varying
proton-electron mass ratio $\mu =m_{p}/m_{e}$, \cite{Barrow2005a,
Scoccola2008}. Scaling arguments have also been used to relate changes in $%
\alpha $ to changes in $\mu $\ using the internal structure of the standard
model, including supersymmetry \cite{Calmet2002}. Typically (in the absence
of unusual cancellations involving the rates of change of $\alpha $, and the
supersymmetry-breaking and grand unification energy scales), they predict
that changes in $\mu $\ at low energies should be about an order of
magnitude greater than those in $\alpha $. However, high-redshift
cosmological bounds on $\mu $\ variation are expected to be weaker than
those from laboratory tests of the equivalence principle \cite{Barrow2005a}.
Systematic investigations of the spectra of cold H$_{2}$ towards quasar
sources have now produced a constraint on $\mu $-variation over cosmological
time scales yielding $\Delta \mu /\mu <1\times 10^{-5}$ at redshifts $z=2-3.5
$, corresponding to look-back times of 10-12 Gyr \cite{Malec2010,
Bagdonaite2014}. \ Radio studies have surpassed optical ones in limiting
changes in $\mu $ at lower redshifts, with $2\sigma $ limits of $\Delta \mu
/\mu <$ few $\times 10^{-7}$ from comparisons between $NH_{3}$ and
rotational lines \cite{kan1,kan2}, and $\Delta \mu /\mu <1.5\pm 1.5\times
10^{-7}$ from multiple methanol lines in a lensing galaxy at $z=0.89$ \cite%
{bag}. Also, at low redshifts, the conjugate satellite OH method is
sensitive to changes in alpha \cite{kan3} at the level of $\Delta \alpha
/\alpha <(-3.1\pm 1.2)\times 10^{-6}$ at $z=0.247$.

Besides producing effects on cosmological scales, the couplings between
light scalar fields and other fields can generate dependencies of coupling
strengths on the local matter density~\cite{sandvik2002,Khoury2004}, or on
local gravitational fields~\cite{Magueijo2002,Flambaum2008}. Such couplings
violate the Einstein equivalence principle \cite{bd,Will2014}. The
gravitational potential at distance $R$ from an object of mass $M$ is
commonly expressed in dimensionless units of $\phi =GM/Rc^{2}$. A number of
studies have been performed using ultrastable lasers and atomic clocks
exploiting the eccentricity of the Earth's orbit~\cite%
{Ferrell2007,Fortier2007,Blatt2008,Shaw2008} causing sinusoidal changes of $%
\Delta \phi =3\times 10^{-10}$. Recently, a spectroscopic study of
Fe\thinspace V and Ni\thinspace V ions in the local environment of the
photosphere of a white dwarf was employed to assess the dependence of $%
\alpha $\ in a strong gravitational field ($\phi =4.9\times 10^{-5}$)~\cite%
{Berengut2013}. Most recently, spectra of molecular hydrogen (H$_{2}$) were
employed to search for any dependence of $\mu $ on gravity. The Lyman
transitions of H$_{2}$, observed with the COS on the Hubble Space Telescope
by Xu et al \cite{Xu2013} towards white dwarf stars GD133 (WD\thinspace 1116$%
+$026) and G29$-$38 (WD\thinspace 2326$+$049) are compared to accurate
laboratory spectra taking into account the high-temperature conditions ($%
T\sim 13\,000$~$K$) of their photospheres to probe possible dependence of $%
\mu $ on a gravitational potential that is $\sim 10^{4}$ times stronger than
its value at the Earth's surface. The spectrum of white dwarf star GD133
yields a $\Delta \mu /\mu $ constraint of $(-2.7\pm 4.7_{\mathrm{stat}}\pm
0.2_{\mathrm{sys}})\times 10^{-5}$ for a local environment with
gravitational potential $\phi \sim 10^{4}\ \phi _{\text{Earth}}$, while that
of G29$-$38 yields $\Delta \mu /\mu =(-5.8\pm 3.8_{\mathrm{stat}}\pm 0.3_{%
\mathrm{sys}})\times 10^{-5}$ for $\phi \sim 2\times 10^{4}$ $\phi _{\text{%
Earth}},$ \cite{WDmols}.

These observational advances lead us to refine our theoretical models and in
what follows we review a new way of viewing the theory of varying $\alpha $
introduced by Bekenstein, Sandvik, Barrow and Magueijo (BSBM) in \cite%
{Bekenstein1982,sandvik2002} and extended in References \cite{lip} and \cite%
{graham}. Following recent work \cite{blackbody} in Section~\ref{diel} we
show how BSBM may be seen as the actions of a dielectric medium permittivity
encoded in scalar field, $\psi $. Specifically, we will see that $\psi $
acts like a relativistic generalisation of a common dielectric or insulator.
Unlike in standard media, $\epsilon $ and $\mu ^{-1}$ obey a relativistic
Klein-Gordon equation sourced by the EM lagrangian, $E^{2}-B^{2}$. The
dielectric analogy suggests a number of extensions of the original BSBM
theory, but if we seek to preserve Lorentz invariance the options are
limited. Regarding the effects of $\psi $ upon electromagnetism, the theory
is fully fixed by Lorentz, parity and gauge invariances, but we may still
change the dynamics of $\psi $. If we wish to extend further the analogy
with a standard dielectric, then it would make more sense to make $\psi $ a
local function of the electromagnetic field. If this function is to be
Lorentz invariant, it can only depend on the scalar $E^{2}-B^{2}$ and the
pseudo-scalar ${\mathbf{E}}\cdot {\mathbf{B}}$, but the latter induces
parity violations (and also permits modifications to Maxwell's equations,
which are beyond the scope of this discussion). The formalism for such 
\textit{local} varying alpha theories is developed in Section~\ref{var1}.

The rest of this paper is devoted to exploring some of the phenomenology of
these theories. As with torsion theories, it is extremely difficult to
constrain the model by local particle physics experiments. In Section~\ref%
{self} we show how the problem of the divergence of the classical
self-energy of a point particle might be removed or ameliorated within these
theories. In Section~\ref{cosmo} we start exploring the cosmological
solutions, particularly as models for the very early universe. We close the
paper with a discussion of the wider implications.

Throughout this paper we shall use Planck units, and employ metrics with
signature is $-,+,+,+$.

\section{Varying-alpha as a relativistic dielectric effect}

\label{diel} In this Section we describe how the scalar field, $\psi $,
which self-consistently drives variations in the electron charge $e\equiv
e_{0}\exp (2\psi )$, and hence in $\alpha $, in BSBM theory, may be
interpreted as a dielectric medium \cite{LL}. The dielectric is linear (in
the sense that $\mathbf{D}$ and $\mathbf{H}$ are proportional to $\mathbf{E}$
and $\mathbf{B}$) and the proportionality constants $\epsilon $ and $\mu
^{-1}$ are isotropic and frequency independent. Unlike in standard media, $%
\epsilon $ and $\mu ^{-1}$ obey a relativistic Klein-Gordon equation sourced
by the EM lagrangian $E^{2}-B^{2}$. Since $\epsilon =1/\mu $, the medium is
non-dispersive, and so it induces no frequency shift or photon production.
The argument was presented in~\cite{blackbody}, and in fact applies to any
varying-alpha theory which preserves relativistic Lorentz invariance and the
gauge principle.

In setting up electromagnetism under a varying alpha there is an
\textquotedblleft ambiguity\textquotedblright\ in the definition of electric
and magnetic fields similar to that found for insulators, where one can use $%
\mathbf{E}$ or $\mathbf{D}$ for the electric field, and $\mathbf{B}$ or $%
\mathbf{H}$ for the magnetic field. In reality both concepts play a role,
with $\mathbf{E}$ and $\mathbf{B}$ convenient for writing the homogeneous
Maxwell equations, and $\mathbf{D}$ and $\mathbf{H}$ better suited for
writing the inhomogeneous equations, even when there are no sources.
Varying-alpha theories may be phrased either in terms of $A_{\mu }$ (as in 
\cite{Bekenstein1982}), or $a_{\mu }$ (as in \cite{sbm1}), with the two
quantities related by: 
\begin{equation}
a_{\mu }=e^{\psi }A_{\mu }\,,
\end{equation}%
where 
\begin{equation}
\psi =\ln \tilde{\epsilon}=\ln \frac{e}{e_{0}}=\frac{1}{2}\ln \alpha .
\end{equation}%
The last expression links $\psi $ to the fine structure \textquotedblleft
constant\textquotedblright , $\alpha $. Here $\tilde{\epsilon}$ corresponds
to the \textquotedblleft $\epsilon $\textquotedblright\ used in \cite%
{Bekenstein1982}, which we stress is \textit{not} the relative permittivity
of the \textquotedblleft medium\textquotedblright ($\epsilon $,\ in our
notation here), as we shall see. Gauge transformations can be performed as 
\begin{equation}
a_{\mu }\rightarrow a_{\mu }+\partial _{\mu }\Lambda
\end{equation}%
or as 
\begin{equation}
A_{\mu }\rightarrow A_{\mu }+\frac{\partial _{\mu }\Lambda }{\tilde{\epsilon}%
}.
\end{equation}%
This fork in the development propagates into the definition of
gauge-invariant field tensors, with \cite{Bekenstein1982} led to make the
natural definition: 
\begin{equation}
F_{\mu \nu }=e^{-\psi }\left[ \partial _{\mu }(e^{\psi }A_{\nu })-\partial
_{\nu }(e^{\psi }A_{\mu })\right] ,  \label{Fdef}
\end{equation}%
and \cite{sbm1} to choose 
\begin{equation}
f_{\mu \nu }=\partial _{\mu }a_{\nu }-\partial _{\nu }a_{\mu }.  \label{fdef}
\end{equation}%
The two are related by 
\begin{equation}
F_{\mu \nu }=e^{-\psi }f_{\mu \nu }.
\end{equation}%
The electromagnetic action, from which the non-homogeneous Maxwell's
equations are derived, can be written in the two forms: 
\begin{equation}
S_{EM}=-\frac{1}{4}\int d^{4}x\,F^{2}=-\frac{1}{4}\int d^{4}x\,e^{-2\psi
}f^{2}.  \label{action}
\end{equation}

In order to study which quantities play the role of $\mathbf{E}$ and $%
\mathbf{B}$ (and so give the equivalent of the Faraday tensor) in~\cite%
{blackbody}, we examined the non-homogeneous Maxwell equations. These are
best written in terms of $f_{\mu \nu }$, in the form of the integrability
condition: 
\begin{equation}
\epsilon ^{\alpha \beta \mu \nu }\partial _{\beta }f_{\mu \nu }=0.
\end{equation}%
This is obviously a necessary condition for (\ref{fdef}), but note that the
same argument cannot be made directly for $F_{\mu \nu }$ (derivatives of $%
\psi $ would appear in the corresponding condition in terms of $F_{\mu \nu }$%
; cf. (\ref{Fdef}) and (\ref{fdef})). Thus, in order to parallel the usual
theory of electrodynamics in media we should associate $\mathbf{E}$ and $%
\mathbf{B}$ (appearing in the inhomogeneous Maxwell equations) with $f_{\mu
\nu }$, with entries in the usual places. With this identification we obtain
the standard inhomogeneous Maxwell equations: 
\begin{eqnarray}
\nabla \cdot {\mathbf{B}} &=&0, \\
\nabla \wedge {\mathbf{E}}+\frac{\partial {\mathbf{B}}}{\partial t} &=&0\;.
\end{eqnarray}%
This was already noted in \cite{Bekenstein1982} (however, the wrong
identification was made in ref.\cite{kraisel}, cf. their Eq.(25)).

In order to find the equivalent of $\mathbf{D}$ and $\mathbf{H}$, in~\cite%
{blackbody} we examined instead the inhomogeneous Maxwell equations. In the
absence of currents these can be written in the two forms: 
\begin{equation}
\partial _{\mu }(e^{-2\psi }f^{\mu \nu })=\partial _{\mu }(e^{-\psi }F^{\mu
\nu })=0,
\end{equation}%
and we see that neither of them leads to the equivalent standard expression
for dielectric media (in both cases extra terms in the derivatives of $\psi $
appear). Therefore, we should define the alternative \textquotedblleft
Maxwell\textquotedblright\ tensor: 
\begin{equation}
\mathcal{F}_{\mu \nu }=e^{-\psi }F_{\mu \nu }=e^{-2\psi }f_{\mu \nu },
\end{equation}%
in terms of which we have 
\begin{equation}
\partial _{\mu }\mathcal{F}^{\mu \nu }=0\,.
\end{equation}%
We should then define $\mathbf{D}$ and $\mathbf{H}$ from the appropriate
entries in $\mathcal{F}_{\mu \nu }$, so as to get: 
\begin{eqnarray}
\nabla \cdot {\mathbf{D}} &=&0,  \label{Deqn} \\
\nabla \wedge {\mathbf{H}}-\frac{\partial {\mathbf{D}}}{\partial t} &=&0\;.
\end{eqnarray}%
With these identifications BSBM becomes equivalent to electromagnetism in
dielectric media with only small adaptions. We have: 
\begin{eqnarray}
\mathbf{D} &=&\epsilon \mathbf{E}=e^{-2\psi }\mathbf{E,} \\
\mathbf{H} &=&\mu ^{-1}\mathbf{B}=e^{-2\psi }\mathbf{B,}
\end{eqnarray}%
and so 
\begin{equation}
\epsilon =\frac{1}{\mu }=e^{-2\psi }.  \label{epsmu}
\end{equation}%
$\mathbf{D}$ and $\mathbf{H}$ are proportional to $\mathbf{E}$ and $\mathbf{B%
}$, and the proportionality constants $\epsilon $ and $\mu ^{-1}$ are
isotropic and frequency-independent.

\section{The permittivity as an algebraic function of the EM field}

\label{var1} The identifications found above suggest an obvious extension of
BSBM. In BSBM, $\psi $ satisfies a driven Klein-Gordon equation, but in
standard electromagnetism the permittivity would be a local function of the
fields $\mathbf{E}$ and $\mathbf{B}$. The only relativistically invariant
such functions take the form: 
\begin{equation}
\psi =\psi (E^{2}-B^{2},\mathbf{E}\cdot \mathbf{B}).  \label{psifunc}
\end{equation}%
In this paper we will focus on the first argument of this function, and
explicitly develop the formalism and applications for this case. Setting up
the formalism for the dependence of $\psi $ on the pseudo-scalar $\mathbf{E}%
\cdot \mathbf{B}$ is a straightforward extension (and we will briefly
present it at the end of this Section). However the phenomenology is
entirely different, involving parity violation effects, and a whole new set
of phenomena. We will defer the study of these theories to a future
publication.

Imposing $\psi=\psi(E^2-B^2) $ can be easily implemented in a Lagrangian
formulation by setting: 
\begin{equation}
S_{EM}=\int d^4 x\, (-\frac{1}{4}e^{-2\psi} f^2 -V(\psi))
\end{equation}
with no kinetic term for $\psi$. Then the Euler-Lagrange equation implies
the algebraic relation: 
\begin{equation}  \label{Vpeqn}
V^{\prime 2\psi}=\frac{f^2}{2}=-(E^2-B^2)
\end{equation}
from which we can deduce $\psi=\psi(E^2-B^2)$. This theory implements
non-linear EM in dielectrics without breaking Lorentz invariance. It differs
from BSBM in many ways, most importantly in that it does not induce a fifth
force or violations of the weak equivalence principle. In fact the theory
has something in common with the basic torsion theories (of the form of the
Einstein-Cartan-Sciama-Kibble type), in that its field is non-propagating.
It is algebraically related to electromagnetic field in this case, rather
than to the spin density, as is the case with torsion.

The Lagrangian of BSBM contains only a kinetic term; instead, the theories
we would like to explore, when recast in the Lagrangian formulation, derive
from Lagrangians which contain only a potential term (and no kinetic term).
Obviously we could add a potential to BSBM, but that would not change its
qualitative nature, since it would still contain propagating degrees of
freedom. We will not consider that possibility here, because we want to
explore the qualitatively different possibility that the permittivity is not
a propagating degree of freedom, but like in normal dielectrics, it is an
algebraic function of the EM field itself.

Note that a \textit{possible} generalization of the formalism to include the
second argument in~(\ref{psifunc}) would follow from 
\begin{equation}
S_{EM}=\int d^{4}x\,\{-\frac{1}{4}e^{-2\psi }f^{2}-\frac{\beta }{4}e^{-2\phi
}f_{\mu \nu }\tilde{f}^{\mu \nu }-V(\psi ,\phi )\}
\end{equation}%
resulting in conditions: 
\begin{eqnarray}
\frac{\partial V}{\partial \psi } &=&-e^{2\psi }(E^{2}-B^{2}) \\
\frac{\partial V}{\partial \phi } &=&-\beta e^{2\phi }({\mathbf{E}}\cdot {%
\mathbf{B}}).
\end{eqnarray}%
In these theories, the Maxwell equations would receive corrections due to
the parity violating term. Other ways to accommodate chirality are possible,
and will be further explored elsewhere.

\section{Removal of the classical self-energy divergence}

\label{self} A suitable choice of $\epsilon$ (or $V(\psi)$) can lead to the
removal or softening of classical divergences, specifically in the particle
electromagnetic self-energy. This was once perceived as a major problem
(citations) and may still interact in non-trivial ways with the more general
issue of quantum divergences. The potential $V$ may be seen as a classical
effective way to describe quantum effects in the self-interaction.
Conversely if we were to postulate it at tree-level, it would affect loop
corrections in non-trivial ways. Regardless of this we now exhibit one
theory where the Coulomb self-energy is finite.

Let us assume that: 
\begin{equation}
V(\psi )=\frac{1}{\ell ^{4}}\left( \frac{1}{2}e^{2\psi }-\psi \right) .
\end{equation}%
Then, solving Eq.~\ref{Vpeqn} leads to two solutions. The branch with the
correct limit when the field is small is: 
\begin{equation}
\epsilon =\frac{1-\sqrt{1-4\ell ^{4}(E^{2}-B^{2})}}{2\ell ^{4}(E^{2}-B^{2})},
\end{equation}%
and we see that this is equivalent to: 
\begin{equation}
{\mathbf{E}}=\frac{\mathbf{D}}{\epsilon }=\frac{\mathbf{D}}{1+\ell ^{4}D^{2}}
\end{equation}%
when ${\mathbf{B}}=0$. Since ${\mathbf{D}}$ satisfies the Poisson equation
(cf. Eq.~(\ref{Deqn})) we have that for a point charge: 
\begin{equation}
\mathbf{D}=\frac{e_{0}}{r^{2}}\mathbf{e_{r}}\,.
\end{equation}%
However now the electric field reaches a maximum and then drops to zero: 
\begin{equation}
\mathbf{E}=\frac{e_{0}}{r^{2}+\frac{\ell ^{4}e_{0}^{2}}{r^{2}}}.
\end{equation}%
Clearly there is no singularity in the energy density around a point charge.
As $r\rightarrow 0$ we find that 
\begin{equation}
E\sim \frac{r^{2}}{l^{4}e_{0}}\,.
\end{equation}%
As explained in~\cite{blackbody}, the electrostatic energy density in a
dielectric (generalized or otherwise) is given by: 
\begin{equation}
\tilde{\rho}_{EM}=\frac{1}{2}\mathbf{E}\cdot \mathbf{D}
\end{equation}%
and so, as $r\rightarrow 0$, the energy density tends to a constant: 
\begin{equation}
\tilde{\rho}_{EM}\rightarrow \frac{1}{2\ell ^{4}}\;.
\end{equation}

We note that there is still a divergence in the energy associated with the
dielectric itself, but this is logarithmic. Indeed the energy associated
with the dielectric can be written as: 
\begin{equation}
\rho _{\psi }=V(\psi )=\frac{1}{\ell ^{4}}\left( \frac{1}{2\epsilon }+\frac{1%
}{2}\log \epsilon \right)
\end{equation}%
and since $\epsilon \rightarrow 1/r^{4}$, the second term diverges only
logarithmically.

\section{Cosmological equations}

\label{cosmo} A closed set of cosmological equations can be found by
appealing to energy conservation. We first review how this is the case in
BSBM, before adapting the argument to the theory proposed in this paper. In
BSBM, the Lagrangian for $\psi $ is: 
\begin{equation}
\mathcal{L}_{\psi }=-\frac{\omega _{B}}{2}\partial _{\mu }\psi \partial
^{\mu }\psi
\end{equation}%
leading to a forced Klein-Gordon equation for $\psi $: 
\begin{equation}
\nabla ^{2}\psi =\frac{2}{\omega _{B}}{\widetilde{\mathcal{L}}}_{EM}\,.
\end{equation}%
where $\widetilde{\mathcal{L}}_{EM}=e^{-2\psi }(E^{2}-B^{2})/2$ (in what
follows we denote by tilded variables those which have factors which are a
function of $\psi $ absorbed into their definitions). Under the assumption
of homogeneity and isotropy, this equation becomes the ODE: 
\begin{equation}
\ddot{\psi}+3\frac{\dot{a}}{a}\dot{\psi}=-\frac{2}{\omega _{B}}\widetilde{%
\mathcal{L}}_{EM}\,.  \label{KG}
\end{equation}%
and this equation can be interpreted as an energy balance equation, with the
driving terms representing energy exchange between $\psi $ and other forms
of matter. Homogeneity and isotropy imply that $\psi $ behaves like a
perfect fluid, and computing the stress-energy tensor reveals: 
\begin{equation}
p_{\psi }=\rho _{\psi }=\omega _{B}\frac{\dot{\psi}^{2}}{2}.  \label{rhopsi}
\end{equation}%
Equation (\ref{KG}) is then equivalent to: 
\begin{equation}
\dot{\rho}_{\psi }+3\frac{\dot{a}}{a}(p_{\psi }+\rho _{\psi })=-2\dot{\psi}%
\widetilde{\mathcal{L}}_{EM}\;.  \label{dotrho}
\end{equation}%
Each component $i$ contributes a term proportional to $\widetilde{\mathcal{L}%
}_{EM_{i}}$ to the right-hand side of (\ref{dotrho}). This should be
balanced by a counter-term with opposite sign in the right hand side of the
the conservation equation for $i$: 
\begin{equation}
\widetilde{\rho }_{i}+3\frac{\dot{a}}{a}(\widetilde{\rho }_{i}+\widetilde{p}%
_{i})=2\dot{\psi}\widetilde{\mathcal{L}}_{EM_{i}}\;.
\end{equation}%
For all $i$ components (including the dark matter) we need equations of
state relating their energy density with their EM Lagrangian content. One
possibility is to define parameters: 
\begin{equation}
\zeta _{i}=\frac{\widetilde{\mathcal{L}}_{i}^{EM}}{\widetilde{\rho }_{i}}\,.
\label{zeta}
\end{equation}%
For radiation $\zeta _{r}=0$, but $\zeta _{m}\neq 0$ for baryonic as well as
for some types of dark matter. We stress that the statement that $\zeta _{i}$
is a constant is part of the model (and such a model is \textit{not} the
model employed for matter in~\cite{sbm1}).

Given these considerations we concluded in~\cite{blackbody} that in BSBM
theory, a full closed set of cosmological equations for a matter and
radiation universe is: 
\begin{eqnarray}
\left( \frac{\dot{a}}{a}\right) ^{2} &=&\frac{1}{3}\left( \widetilde{\rho }%
_{m}+\widetilde{\rho }_{r}+\rho _{\psi }\right) ,  \label{fried0} \\
\dot{\widetilde{\rho }}_{r}+4\frac{\dot{a}}{a}\widetilde{\rho }_{r} &=&0, \\
\dot{\widetilde{\rho }}_{m}+3\frac{\dot{a}}{a}\widetilde{\rho }_{m} &=&2\dot{%
\psi}\zeta _{m}\widetilde{\rho }_{m}, \\
\dot{\rho}_{\psi }+6\frac{\dot{a}}{a}\rho _{\psi } &=&-2\dot{\psi}\zeta _{m}%
\widetilde{\rho }_{m}.
\end{eqnarray}%
Here, $\dot{\psi}$ on the right-hand side of the last two equations is to be
written as 
\begin{equation}
\dot{\psi}=\sqrt{\frac{2\rho _{\psi }}{\omega _{B}}}
\end{equation}%
(where we used (\ref{rhopsi})) to form a closed system.

A similar construction can be set up for the theory proposed in this paper,
with (\ref{KG}) replaced by (\ref{Vpeqn}), or, using the notation in this
Section: 
\begin{equation}
V^{\prime }=-2{\widetilde{\mathcal{L}}}_{EM}.  \label{vpeq1}
\end{equation}%
In view of the considerations leading to (\ref{zeta}), we can rewrite this
as: 
\begin{equation}
V^{\prime }=-2\zeta _{m}\widetilde{\rho }_{m}\,.  \label{Vpeqn1}
\end{equation}%
Assuming homogeneity and isotropy, the $\psi $ fluid is now made up of pure
potential energy, instead of kinetic energy (as was the case for BSBM), so,
instead of (\ref{rhopsi}), we have 
\begin{equation}
-p_{\psi }=\rho _{\psi }=V(\psi ).  \label{rhopv}
\end{equation}%
The conservation equation for the $\psi $-fluid therefore reads: 
\begin{equation}
\dot{\rho}_{\psi }+3\frac{\dot{a}}{a}(\rho _{\psi }+p_{\psi })=\dot{\rho}%
_{\psi }=\dot{\psi}V^{\prime }=-2\dot{\psi}\widetilde{\mathcal{L}}_{EM},
\end{equation}%
where in the last identity we have used (\ref{vpeq1}), and in the first two
we have used (\ref{rhopv}). We therefore obtain an equation identical to (%
\ref{dotrho}) but with a different equation of state for the $\psi $ fluid.
It can be further expressed as: 
\begin{equation}
\dot{\rho}_{\psi }=-2\dot{\psi}\zeta _{m}{\widetilde{\rho }}_{m}
\end{equation}%
As a result, we know that the right-hand side (RHS) of this equation should
appear with a reversed sign as a source term to the the matter conservation
equation. The cosmological equations for a radiation-matter universe are
therefore: 
\begin{eqnarray}
\left( \frac{\dot{a}}{a}\right) ^{2} &=&\frac{1}{3}\left( \widetilde{\rho }%
_{m}+\widetilde{\rho }_{r}+\rho _{\psi }\right)  \label{fried} \\
\dot{\widetilde{\rho }}_{r}+4\frac{\dot{a}}{a}\widetilde{\rho }_{r} &=&0 \\
\dot{\widetilde{\rho }}_{m}+3\frac{\dot{a}}{a}\widetilde{\rho }_{m} &=&2\dot{%
\psi}\zeta _{m}\widetilde{\rho }_{m} \\
\dot{\rho}_{\psi } &=&-2\dot{\psi}\zeta _{m}\widetilde{\rho }_{m}.
\end{eqnarray}%
This system looks formally identical to the equations obtained for BSBM
(once one accounts for the different equation of state for $\psi $), but in
fact the system is entirely different and, in fact in this form, it does not
constitute a closed set of ODEs. Given that we have (\ref{rhopv}) instead of
(\ref{rhopsi}) we cannot write $\dot{\psi}$ on the RHS of the last two
equations as a function of $\rho _{\psi }$. Instead, we should use 
(\ref{Vpeqn1}) to write $\psi $ as a function of $\widetilde{\rho }_{m}$. We
can then write $\dot{\psi}$ as: 
\begin{equation}
\dot{\psi}=\frac{d\psi }{d\widetilde{\rho }_{m}}\dot{\rho}_{m}.
\end{equation}%
Given that $\psi =\psi (\widetilde{\rho }_{m}),$ we can also eliminate $\rho
_{\psi }$ from the equations, since $\rho _{\psi }$ in the Friedman equation
can be written as $\rho _{\psi }=V(\psi )=V(\psi (\widetilde{\rho }_{m}))$.
These two steps allow for a rearrangement of the equations into a closed
system: 
\begin{eqnarray}
\left( \frac{\dot{a}}{a}\right) ^{2}=\frac{1}{3}\left( \widetilde{\rho }_{m}+%
\widetilde{\rho }_{r}+V(\widetilde{\rho }_{m})\right) , &&  \label{friedH} \\
\dot{\widetilde{\rho }}_{r}+4\frac{\dot{a}}{a}\widetilde{\rho }_{r} &=&0, \\
\dot{\widetilde{\rho }}_{m}{\left( 1-2\zeta _{m}\widetilde{\rho }_{m}\frac{%
d\psi }{d\widetilde{\rho }_{m}}\right) }+3\frac{\dot{a}}{a}\widetilde{\rho }%
_{m} &=&0.  \label{consH}
\end{eqnarray}%
As our worked examples will now show, in practice these steps are always
folded into finding a solution to the system.

\section{Some examples}

We now consider some cosmological application of this theory. We construct
models for the early universe, assuming for simplicity a single form of
matter ($i=1$) with constant $\zeta \neq 0$ and a general equation of state $%
p/\rho =w$, with $w$ constant. It is curious that the simplest choices of
potential lead to interesting solutions, namely bouncing, loitering and
inflationary dynamics.

\subsection{Quadratic potential: bouncing models}

\label{massV} Let us assume a quadratic potential, leaving the sign
undefined for the time being: 
\begin{equation}
V(\psi )=\pm \frac{1}{2}M^{4}\psi ^{2}.
\end{equation}%
The Euler-Lagrange equation leads to: 
\begin{equation}
V^{\prime 4}\psi =-2\zeta \widetilde{\rho },
\end{equation}%
and so we learn that this a model in which the early universe is filled with
a dielectric rendering the electric charge exponentially dependent on the
density: 
\begin{equation}
e=e_{0}\exp {\psi }=e_{0}\exp {\frac{\mp 2\zeta \widetilde{\rho }}{M^{4}}}.
\end{equation}%
Furthermore, we have: 
\begin{eqnarray}
\psi &=&\mp 2\frac{\zeta }{M^{4}}\widetilde{\rho } \\
V(\widetilde{\rho }) &=&\pm 2\frac{\zeta ^{2}}{M^{4}}\widetilde{\rho }^{2}.
\end{eqnarray}%
Therefore, following the procedure outlined at the end of last Section, we
find that the Friedmann equation (see Eq.~(\ref{friedH})) resembles that
obtained in brane-world cosmology when the minus sign is picked: 
\begin{equation}
\left( \frac{\dot{a}}{a}\right) ^{2}=\frac{1}{3}\left( \widetilde{\rho }\pm 2%
\frac{\zeta ^{2}}{M^{4}}\widetilde{\rho }^{2}\right) .
\end{equation}%
However, this theory is very different because the Friedmann equation is
supplemented by a modified conservation equation (cf. Eq.~(\ref{consH})): 
\begin{equation}
\dot{\widetilde{\rho }}{\left( 1\pm 4\frac{\zeta ^{2}}{M^{4}}\widetilde{\rho 
}\right) }+3(1+w)\frac{\dot{a}}{a}\widetilde{\rho }=0,
\end{equation}%
and therefore the dynamics is potentially very different.

It is easy to prove that the theory still leads to bouncing behaviour. The
conservation equation integrates to: 
\begin{equation}
\ln \widetilde{\rho }\pm \frac{4\zeta ^{2}}{M^{4}}\widetilde{\rho }=-3\ln a
\end{equation}%
or: 
\begin{equation}
\widetilde{\rho }\exp \left( {\pm 4\frac{\zeta ^{2}}{M^{4}}\widetilde{\rho }}%
\right) \propto \frac{1}{a^{3(1+w)}}
\end{equation}%
If we choose the minus sign for the potential this model leads to a bounce
when the density reaches the maximum: 
\begin{equation}
\widetilde{\rho }_{max}=\frac{M^{4}}{2\zeta ^{2}}.
\end{equation}%
We would also expect the $\rho ^{2}$ term to dominate the effects of simple
shear anisotropies on approach to the expansion minimum and create a
quasi-isotropic bounce whenever $w>0$.

\subsection{Exponential potential: Inflation and loitering}

\label{expV} Another interesting case is that with an exponential potential: 
\begin{equation}
V(\psi )=V_{0}e^{-\lambda \psi }.
\end{equation}%
Then: 
\begin{equation}
V^{\prime }=-\lambda V_{0}e^{-\lambda \psi }=-2\zeta \widetilde{\rho },
\end{equation}%
so that 
\begin{eqnarray}
\psi &=&-\frac{1}{\lambda }\ln \widetilde{\rho }+C \\
V &=&\frac{2\zeta }{\lambda }\widetilde{\rho }.
\end{eqnarray}%
We therefore now have a model in which the early universe behaves like a
dielectric for which the electric charge is a power-law of the energy
density 
\begin{equation}
e=e_{0}\exp {\psi }\propto {\widetilde{\rho }}^{-\frac{1}{\lambda }}.
\end{equation}%
Inserting the solution into Eq.~(\ref{consH}) gives us the modified
conservation equation 
\begin{equation}
\dot{\widetilde{\rho }}{\left( 1+B\right) }+3(1+w)\frac{\dot{a}}{a}%
\widetilde{\rho }=0
\end{equation}%
with $B=2\zeta /\lambda $. Thus, regarding matter evolution, there is an
effective shift in the equation of state: 
\begin{equation}
w\rightarrow w_{eff}=\frac{w-B}{1+B}.
\end{equation}%
We see that as $B\rightarrow \infty $ we get $w_{eff}\rightarrow -1$, i.e.
inflation. If $w=0$, for $B>1/2$ we have $w_{eff}<-1/3$. This shift in the
equation of state transfers into the expected change in the expansion rate,
because $V\propto \widetilde{\rho }$, 
and so the Friedmann equation reads: 
\begin{equation}
\left( \frac{\dot{a}}{a}\right) ^{2}=\frac{1}{3}\left( 1+\frac{2\zeta }{%
\lambda }\right) \widetilde{\rho }.
\end{equation}%
The only effect in this equation is an effective shift in the gravitational
constant 
\begin{equation}
G\rightarrow G_{eff}=G(1+B).
\end{equation}%
We can therefore generate acceleration (certainly an early-time,
inflationary one) with this model.

The extreme case $B=-1$, corresponding to 
\begin{equation}
\lambda =-2\zeta ,
\end{equation}%
is interesting in that gravity seemingly switches off (since $G_{eff}=0$). A
static universe is then possible. Close to this value we may induce
loitering.

\section{Conclusions}

In this paper we have examined an alternative to BSBM varying-alpha theory
inspired by conventional dielectrics. The properties of BSBM theory may be
understood from the standard electrodynamics of dielectrics with suitable
definitions for the fields $\mathbf{E}$ and $\mathbf{B}$ and the
displacements $\mathbf{D}$ and $\mathbf{H}$ (associated with the Faraday and
Maxwell tensors, respectively). We must, however be prepared to regard the
\textquotedblleft vacuum\textquotedblright\ as a dielectric medium with
unusual properties: Lorentz invariant and with $\epsilon =1/\mu $ satisfying
a driven Klein-Gordon equation. Conventional dielectrics, however, are not
dynamical but, rather, have properties which are local functions of the EM
field. We explored this possibility in this paper without letting go of
Lorentz invariance. The result is a theory with \textit{formal} analogies to
torsion in the Einstein-Cartan-Sciama-Kibble theory of gravity. If the
connection and metric are to be seen as independent degrees of freedom, then
torsion is inevitable, but it appears as a non-dynamical degree of freedom,
algebraically related to the spin-density. Here we find an algebraic
relation between $\epsilon =1/\mu $ and the EM Lagrangian. In both cases it
is extremely difficult to constrain the ensuing theory.

Nonetheless the theory leads to interesting results. At a very fundamental
level it could change the outlook on the problem of the divergence of the
self-energy of particles, as we have shown in Section~\ref{self}. It also
leads to interesting early universe cosmologies, with simple choices for the
potential inducing bounces, loitering and acceleration. The dynamics is
fundamentally different from that found in BSBM, because the conservation
equation must be modified so as to obtain a closed system of governing
equations. The late-time implications are likely to be less dramatic. With a
quadratic potential the effects are suppressed by factors $O(\rho
_{m}/M^{4}) $ and so are naturally small (one can easily adapt the
calculations in Section~\ref{massV} to a late-time matter universe). With an
exponential potential we obtain tracking solutions, just as with
quintessence, but these induce a shift in the equation of state. We should
therefore either ensure that for dark matter the constant $B$ is small, or
else induce a feature in the potential (i.e. a local minimum) taking the
field away from rolling behaviour at late-times.

Further extensions of this theory are possible. We may regard BSBM as a
\textquotedblleft purely kinetic\textquotedblright\ varying-alpha theory (in
which the Lagrangian for $\psi $ only has a kinetic term). By contrast, the
theory proposed here is endowed with a Lagrangian with only a potential
term, leading to an algebraic relation between $\alpha $ and EM field. We
could of course have both a kinetic and a potential term, but such a theory
would ontologically be more like a BSBM theory, where $\psi $ is a truly
dynamical field. Quantitative differences would arise, and one suspects that
scaling solutions similar to those found in the quintessence scenario would
exist. A more dramatic possible generalisation of these theories would arise
from giving up Lorentz symmetry. Then a much larger class of theories can be
written down, with a richer phenomenology. We are currently exploring this
possibility.

\section*{Acknowledgments}

JM and JDB acknowledge STFC consolidated grant support. JM was also
supported by the Leverhulme Trust and the John Templeton Foundation. We
thank N. Kanekar for helpful comments.

\end{document}